\documentclass[review]{elsarticle}

\usepackage{amsfonts}
\usepackage{lineno,hyperref}
\usepackage{amsmath}
\usepackage{natbib}
\usepackage{array}
\newcolumntype{F}[1]{%
    >{\raggedright\arraybackslash\hspace{0pt}}p{#1}}%
\newcolumntype{T}[1]{%
    >{\centering\arraybackslash\hspace{0pt}}p{#1}}%
\modulolinenumbers[5]

\newcommand{\boldX}{{\bf X}}

\journal{a Journal}

\biboptions{authoryear}

\begin{document}

\begin{frontmatter}

\title{{Forecasting number of corner kicks taken in association football using compound Poisson distribution}}


\author[add1]{Stan Yip\corref{cor1}}
  \ead{gstanyip@connect.polyu.hk}
  \author[add1]{Yinghong Zou\corref{au2}}\ead{yinghong.zou@connect.polyu.hk}
  \author[add2]{Ronald Tsz Hin Hung\corref{au3}}\ead{19481217@life.hkbu.edu.hk}
  \author[add1]{Ka Fai Cedric Yiu\corref{au4}}\ead{cedric.yiu@polyu.edu.hk}

  \cortext[cor1]{The Hong Kong Polytechnic University, Hung Hom}
  \address[add1]{Department of Applied Mathematics, The Hong Kong Polytechnic University, Hong Kong}
  \address[add2]{Department of Economics, Hong Kong Baptist University, Hong Kong}
 

\begin{abstract}
    This article presents a holistic compound Poisson regression model framework to forecast number of corner kicks taken in association football. Corner kick taken events are often decisive in the match outcome and inherently arrive in batch with serial clustering pattern. Providing parameter estimates with intuitive interpretation, a class of compound Poisson regression including a Bayesian implementation of geometric-Poisson distribution is introduced. With a varying shape parameter, the corner counts serial correlation between matches is handled naturally within the Bayesian model. In this study, information elicited from cross-market betting odds was used to improve the model predictability. Margin application methods to adjust market inefficiency in raw odds is also discussed.
\end{abstract}


\end{frontmatter}


\section{Introduction}
Association football is the most popular sport in the world with billions of followers. The sport drives significant contribution to the economy from the global television revenues \citep{roberts2016value}, tourism income from extensive travels \citep{allmers2009economic} and sponsorship from betting operators \citep{deutscher2019demand}. Thanks to the rapid development of broadcasting and big data technology, a considerable amount of information from different sources allows forecasting the sport outcomes more accessible. 

Forecasting the likelihood of football match outcomes is a timely research topic within the statistics and econometrics community in the past decades. There are at least 40,000 articles in JSTOR and 3,700 entries in Econ Lit that refer to sports events modelling and related factors \citep{stekler2010issues}. Quantifying uncertainty of the association football outcomes first attracts the attention from statisticians.  The seminal paper by \cite{maher1982modelling} utilises a bivariate Poisson distribution to capture teams' inherent attacking and defensive strengths factors. \cite{dixon1997modelling} elaborate Maher's work to exploit the betting market inefficiency. \cite{koopman2019forecasting} provide a dynamic forecasting model with time-varying coefficients that generates a significant positive return over the bookmaker's odds. \cite{angelini2017parx} use a Poisson autoregression with exogenous covariates (PARX) developed by econometricians to model the football matches outcomes. \cite{ley2019ranking} provide 10 different methods to rank football teams based on their historical performance. \cite{baboota2019predictive} adopt a gradient boosting to determine English Premier League results. A binary dynamic time-series model is introduced in \cite{mattera2021forecasting} for forecasting two-way football outcomes such as red cards occurrence, total goals over/under (O/U) and goal/no goal events. Hierarchical Bayes framework is often used in modelling football which can naturally handle the relations between unobserved variables and componentise a model into different parts. \cite{rue2000prediction} perform a Bayesian predictive and retrospective study in the evolution of football team strengths by incorporating dynamic attacking and defensive factors. \cite{baio2010bayesian} develop a Bayesian model to formulate home advantage and latent team effects. In \cite{karlis2009bayesian}, goal supremacy is adequately described with a Bayesian Skellam regression model.

Several attempts have been made to extract information from the pre-match betting market and expert pundits. \cite{clarke2017adjusting} study the forecasting ability after adjusting bookmaker’s odds to allow for overround. \cite{cain2000favourite} study a high odds overvaluation phenomenon termed favourite-longshot bias in the UK football betting market. \cite{deschamps2007efficiency} find a negative favourite-longshot bias on draw odds in the home-away-draw (HAD) market. Pundits and experts information are often considered by the general public. However, \cite{forrest2005odds} discover that subjective forecasting is inferior to model-based forecast. Forecasts from statistical model often contain the information that the betting market not taken into account \citep{pope1989information,goddard2004forecasting, dixon2004value}. \cite{sung2007comparing} illustrate that combining probabilities from statistical models and instant market odds with a simple statistical model improves the accuracy probabilistic forecast.

Apart from the simple HAD prediction, corners kicks are also of both sport analytics and betting interest. Corner kicks are considered to be a reasonable goal-scoring opportunity. \cite{pulling2013defending} study the tactical behaviour when defending corner kicks. \cite{casal2015analysis} provide a thorough analysis on the factors to increase the likelihood of shot-on-goal from corner kicks. \cite{fitt2006valuation} give a valuation framework on the most commonly traded football spread bets including the total number of corners betting option. Corner counts market is one of the most popular sidebet markets offered by sports bookmakers. A more advanced understanding on the statistical properties of the corner counts distribution helps the bookmakers to manage risk through setting the odds more adequately. Number of corners taken is also an important match statistics that never available prior to the match. Along with other match statistics such as number of shots and number of interceptions throughout the game, if these match statistics can be forecasted prior to the match start, it contributes more information for predicting match outcomes \citep{wheatcroft2021forecasting}.

Data provenance is a crucial component of an accurate forecasting system. Data collected in football has lately become a concern due to novel sensor modalities, with potentially high commercial and research interest \citep{stein2017make}. \cite{hutchins2016tales} discuss that the infrastructure and resources necessary to generate real-time data are contributing to rising disparities between top 'data-rich' sports and comparatively poor 'data-poor' sports.

The remainder of this article is organised as follows. Section 2 discusses the data and the information extraction issues. Section 3 introduces the family of compound Poisson distributions and model development. Section 4 implements the algorithm and performs betting simulation. A detailed discussion of the model implications and future work is provided in Section 5.

\section{Data acquisition and information extraction}

\subsection{Data acquisition}

Football match information were collected from the Hong Kong Jockey Club (HKJC) website (www.hkjc.com) 
which includes 20,190 matches with corner over-under market offered from 1st August 2016 to 10 June 2021 across 90 leagues and cup games. Pre-match HAD, total goals scored over-under and  total corner over-under market odds are recorded. On the other hand, post-match information including scorelines and number of corners are also obtained. The HKJC set the market odds at least a day before the match starts. All the prices collected are the odds just before the match start. 

The integrated dataset used in this study (Table \ref{tab:cornerdatalist}) also consists of an alternative data source with extra match information from seven elite leagues (i.e.: English Premier League, Scottish Premier League, German Bundesliga, Italian Serie A, Spanish La Liga, French Ligue 1 and Dutch Eredivisie) including number of shots, number of shot-on-goals, number of fouls and number of cards given and number of the corner kicks taken by each team. These can be downloaded from \url{www.football-data.co.uk}. A mild correlation $(9.68\%/17.37\%)$ between the number of shots/shot-on-goals and the number of corner counts is found.

In order to estimate parameters more accurately and construct useful covariates from bookmaker odds in the model, utilising a proper overdispersed distribution due to clustered counts (not due to excess of zeros and attacking/defensive tactics) is crucial in the information extraction process which will be discussed in the following subsections.

\begin{table}[hbt!]
\caption{Top 15 competitions with total corner over-under market odds offered by HKJC from 1st August 2016 to 10 June 2021}
\begin{tabular}{lF{0.28\textwidth}F{0.28\textwidth}}
\hline
Competitions & Number of corner markets offered & Extra match statistics available\\\hline
Italian Serie A & 1863 & \text{Yes}\\ 
English Premier League & 1849 & \text{Yes}\\ 
Spanish La Liga & 1784 & \text{Yes}\\ 
USA Major League Soccer & 1766 & \text{No}\\ 
German 2. Bundesliga & 1517 & \text{No}\\ 
German Bundesliga & 1503 & \text{Yes}\\ 
Japanese J-League & 1494 & \text{No}\\ 
French Ligue 1 & 1066 & \text{Yes}\\ 
Europa Cup & 942 & \text{No}\\ 
Australian A-League & 722 & \text{No}\\ 
European Champions League & 675 & \text{No}\\ 
Russian Premier League & 475 & \text{No}\\ 
Japanese J-League 2 & 305 & \text{No}\\ 
Asian Champions League & 300 & \text{No}\\ 
Korean K-league & 282 & \text{No}\\ 
Other 75 leagues and competitions & 3647 & \text{No}\\
\textbf{Total} & \textbf{20190}\\
\hline
\end{tabular}
\label{tab:cornerdatalist}
\end{table}

\subsection{Overdispersed behaviour of the total goals scored and corner kicks taken}

Overdispersion in Poisson models occurs when the response variance exceeds the mean \citep{hilbe2011negative}. Apart from attributing to the excess of zero counts in the statistical distribution \citep{lambert1992zero} or the dependency to another variable \citep{kocherlakota1988compounded,kocherlakota2001regression,i2009priori,sengupta2016bivariate}, it also arises from serial correlation between events or counts of event are clustered that violates the distributional assumption of Poisson random variable. Weak overdispersion occurs in association football total goals scored \citep{rodriguez2017regression} whilst this statistical property can also be found in many other sports such as basketball, baseball, hockey \citep{higgs2021bayesian} in varying degrees. Unlike some American sports that the scores depending on numbers of factor brought out extra variation \citep{pollard1973collegiate}, the nature of the overdispersion in football total goals scored is inherently different. \cite{maher1982modelling} asserts that Poisson distribution is a better model than the overdispersed negative binomial distribution due to its close-to-one variance-to-mean ratio. \cite{dixon1997modelling} introduce a low scoring modification on a bivariate Poisson to handle teams' changing tactics when two teams are one goal apart. Contrast to the overdispersed behaviour of the total goals scored, modelling overdispersed corner kicks taken counts is a relatively under-explored area that no known literature has ever discussed it.

\begin{figure}[hbt!]
  \noindent\includegraphics[width=30pc,angle=0]{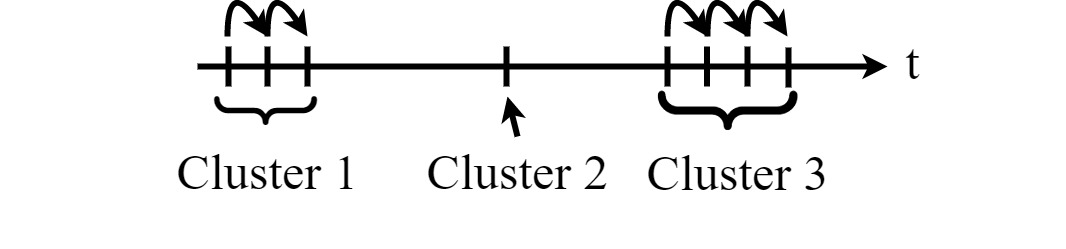}\\
  \caption{Schematic diagram to illustrate how corner events spawned from its ``parent'' event.}
  \label{fig:batch}
\end{figure}

In the HKJC dataset recorded matches from 2016-2021, the sample variance-to-mean ratio for the number of total goals scored has a value of $1.0396$ which agrees with the suggestion in \cite{maher1982modelling} that the scoreline exhibits a Poisson distribution. On the other hand, the ratio for the number of corner kicks taken is $1.1856$ which indicates a fairly significant overdispersion. The corner kick clusters are arisen from ``parent'' corners or offensive duels which produces one or more ``offsprings'' through multiple clearances off the goal line that last touched a player of the defending team (for schematic illustration, see Fig. \ref{fig:batch}). A cluster of corner kicks is understood as a few corners occurring in a short time span. In other words, corners from open play might be a Poisson, but the number of total corners kicks taken rather follows an overdispersed distribution. Failing to recognise this type of temporal clustering leads to an underestimation of number of corner kicks taken.

\subsection{Bookmaker odds as forecasters and margin removal methods}
Betting odds issued by bookmakers reflect implicit probabilities about sporting outcomes \citep{stekler2010issues,leitner2010forecasting,kovalchik2016searching,clarke2017adjusting}. Some empirical evidences show that the implied probabilities from betting odds provide moderate accuracy and outperform tipsters prediction \citep{spann2009sports,stekler2010issues, reade2014information}. The reciprocals of the bookmaker odds are understood as the bookmaker's probabilistic belief. These values sum to more than one. The sum of these probabilities $\pi$ is known as the margin of the book which determines how large the edge toward to the bookmaker. Its inverse $1/\pi$ is the payout rate which indicates the expected value from the bettor point of view. The method to normalise this probabilistic belief is called margin removal method.

It is found that the market often overvalues low probability events. \cite{cain2000favourite} examine that there is a favourite-longshot bias in the football fixed-odds betting market. One explanation is that there are well-informed insiders in the market and the bookmakers maximises their profits through applying less percentage of margin to a particular selection in the book \citep{shin1992, shin1993measuring}. Another popular interpretation is that the bias come from amateur bettors' risk-loving behaviour who overvalue longshot selections. 

\subsubsection{Multiplicative margin removal}
The multiplicative margin removal method, also known as basic method, normalise the inverse odds proportionally and divided by its booksum $\pi$. Most studies has widely adopt this method due to simplicity \citep{vstrumbelj2014determining}. Let $\mathbf{o}=\left(o_1,o_2,\ldots,o_l\right)$ be the quoted decimal odds for a football match with $l\ge 2$ possible outcomes and the inverse odds are $\boldsymbol{\Pi}=\left(\pi_1,\pi_2,\ldots,\pi_l\right)$. Four popular margin removal methods are considered in this article. The implied probability derived from the multiplicative margin removal method is:
$$
    p_i=\pi_i/\pi,
$$
where $\pi = \sum_i^{l}\pi_i$ is the booksum.

\subsubsection{Odds ratio margin removal}
Implemented in the R package \texttt{implied} \citep{implied}, the odds ratio margin removal method fixes the ratio between the implied probability and the raw probability with margin $\mathbf{\pi}$ on all possible outcomes. The odds ratio (OR) is defined as:
$$
    \text{OR} = \frac{p_i (1-\pi_i)}{\pi_i (1-p_i)},
$$
where OR is selected so that $\sum_{i}^{l}p_i = 1$. The implied probability can be calculated with the following equation:
$$
    p_i=\frac{\pi_i}{\text{OR} + \pi_i - \left(\text{OR} \times \pi_i\right)}.
$$
In spite of its mathematical elegance and the ability to capture the favourite-longshot bias, this method is relatively unknown to academic community.

\subsubsection{Shin's method}
Shin's method handles the information asymmetry for bookmakers as less-informed price-setters who face a group of insiders with superior information. \cite{vstrumbelj2014determining} finds that the implied probabilities derived from the betting odds using Shin's method are more accurate than those derived from other simple methods. The formula of taking out margin from the raw probability is given in \cite{jullien1994measuring}:
$$
    p_i=\frac{\sqrt{z^2+4\left(1-z\right)\frac{\pi_i^2}{\pi}}-z}{2\left(1-z\right)},
$$
where $z$ can be numerically estimated by an iterative method starting at $z_0=0$:
$$
    z=\frac{\sum_{i=1}^l\sqrt{z^2+4(1-z)\frac{\pi_i^2}{\pi}}-2}{l-2}.
$$
In the special case of $l=2$, \cite{clarke2017adjusting} show that it is equivalence to a rarely used additive method that does not guarantee a positive probability for $l>2$ and has a tractable analytic solution:
$$
z = \frac{(\pi_+-1)(\pi_-^2-\pi_+)}{\pi_+(\pi_-^2-1)},
$$
where $\pi_+=\pi_1+\pi_2$ and $\pi_-=\pi_1-\pi_2$.

\subsubsection{Power margin removal}
The power method is originally described in \cite{vovk2009prediction} and further developed in \cite{clarke2016adjusting} and \cite{clarke2017adjusting}. The power parameter $k$ scales the inverse odds to the implied probabilities:
$$
    p_i = \pi_i^{1/k},
$$
where the parameter $k$ is chosen that $\sum_{i}^{l} p_i = 1$. The power method outperforms other marginal removal methods in the ATP men's singles tennis two-way market. It is thought that the method is commonly used by bookmakers through empirical evidence \citep{clarke2017adjusting}.

\subsection{Cross-market information}
Market odds from HAD and O/U provides information of how the market view the intensities of goals in a football match. It is generally understood that the expected number of total goals scored and the relative strength between home and away teams play a strong role in explaining number of corners. In order to incorporate such an information, implied expected number of goal scores by both teams can be extracted from the double independent Poisson model proposed by \citep{maher1982modelling}:
$$
           f_{X_1,X_2}(x_1,x_2 | \lambda_1, \lambda_2) =  \frac{\lambda_1^{x_1} e^{-\lambda_1}}{x_1!} \times \frac{\lambda_2^{x_2} e^{-\lambda_2}}{x_2!}.
$$
The implied expected total goals scored $\textbf{TG}_i$ and home team's relative goal supremacy $\textbf{SUP}_i$ of match $i$ can be reparameterised from the implied values of $\hat{\lambda}_1$ and $\hat{\lambda}_2$:
$$
    \begin{array}{lll}
        \textbf{TG}_i & = &  \hat{\lambda}_1 + \hat{\lambda}_2,\\
        \textbf{SUP}_i & = & \hat{\lambda}_1 - \hat{\lambda}_2.
    \end{array}
    \label{eqn:maher}
$$

A variant of the limited memory Broyden–Fletcher–Goldfarb–Shanno algorithm (L-BFGS-B; Bryd et al., 1995\nocite{byrd1995limited}) is used to optimise the square loss function of $\hat{\lambda}_1$ and $\hat{\lambda}_2$ for each match with implied probabilities from HAD and total goals scored O/U odds after removing bias. The implied parameters $\textbf{TG}_i$ and $\textbf{SUP}_i$ from the market odds are used to reflect how the market account for the match expected total goals scored and goal supremacy between two teams. Therefore, the L-BFGS-B algorithm is used to minimise the following loss function:
$$
\left(p_{H}-p'_{H}\right)^2 + \left(p_{D}-p'_{D}\right)^2 + \left(p_{L}-p'_{L}\right)^2,
$$
where $p_{H}$, $p_{D}$, $p_{L}$ are the implied probabilities from the match HAD home, HAD draw, O/U under odds respectively. The model HAD home, HAD draw and O/U under probabilities $p'_{H}$, $p'_{D}$, $p'_{L}$ are derived from the Equation (\ref{eqn:maher}). 

\subsection{Competition-dependent factor and team-level historical records}

Over 90 leagues or cup competitions being offered the total number of corners taken O/U market by the HKJC during the period 2016-2021. Table \ref{tab:cornerlist} shows the average corner counts in some competitions are higher and inherently more volatile with a higher variance-to-mean ratio which is estimated by maximising the log-likelihood of a negative binomial generalised linear model. Target encoding techniques transform categorical variables to a numerical covariate by a statistic computed using the predictors. The following equation generalises the target encoding technique proposed by \cite{micci2001preprocessing} to replace the categorical variable by a numerical value $\hat{x}^i_k$ of match $i$ that belong to competition $k$:
$$
           \hat{x}^i_k = \frac{n^{i}_{k}\cdot \hat{\theta_k}+\hat{\theta}m}{n^{i}_{k}+m},
           \label{eqn:te}
$$
where $n^i_k$ is the number of matches in competition $k$, $\hat{\theta}$ is the statistic of interest such as logarithm of average corner counts and maximum likelihood estimator of the shape parameter, $\hat{\theta_k}$ is the statistic of interest on competition $k$ and $m$ is a hyperparameter that set to a reasonable sample size that makes an estimator. Although target leakage problem is a common cause of overfitting in many machine learning algorithms \citep{zhang2013domain,prokhorenkova2017catboost}, this problem is negligible in this study since the covariates encoded $\hat{x}^i_k$ via the target encoding method is obtained from a pre-training period of 2016-2018 only which is not a part of the data in the fitted model.

\begin{table}[hbt!]
\caption{Average number of corner kicks taken and estimated variance-to-mean ratio $D$ by competition using the \texttt{glm.nb} function in R  for the leagues with sample size greater than 500 during the study period of 2016-2021}
\begin{tabular}{lrr}
\hline
Competitions & Average number of corners & $\hat{D}$ \\\hline
Australian A-League & 11.04 & 1.20\\
Italian Serie A & 10.43 & 1.25\\
English Premier League & 10.38 & 1.14\\
USA Major League Soccer & 10.18 & 1.12\\
German 2. Bundesliga & 10.00 & 1.16\\
UEFA Champions League & 9.73 & 1.30\\
Japanese J-League & 9.70 & 1.11\\
French Ligue 1 & 9.65 & 1.13\\
German Bundesliga & 9.58 & 1.19\\
UEFA Europa League & 9.40 & 1.23\\
Spanish La Liga & 9.39 & 1.10\\

\hline
\end{tabular}
\label{tab:cornerlist}
\end{table}


\section{Model development}

\subsection{Discrete compound Poisson distribution}

The discrete compound Poisson (DCP) distribution is a broad family of distributions widely used in modelling contagious disease attacks, repeated accidents \citep{greenwood1920inquiry,feller}, bacteria spawn \citep{neyman1939new} and batch arrival counts \citep{adelson1966compound}. It is a reasonable model to consider a Poisson process for each cluster of counts and it also captures the intensity varied from cluster to cluster. With Poisson distribution as its special case, it enjoys a great flexibility for overdispersed count data. Suppose the number of clusters $N$ whose distribution is given as:
$$N \sim \textbf{Poisson}(\lambda).$$
A DCP random variable $Y$ is defined by the form:
\begin{equation}
Y = \sum_{i=1}^{N}X_i,
\label{eqn:dcp}
\end{equation}
where $X_1,X_2,\ldots,X_N$ are independent identically distributed discrete random variables. A random variable $Y$ is said to be a DCP random variable \citep{wimmer1996multiple} when its probability generating function (pgf) $G(s)$ has a form:
\begin{equation}
G_Y(s) = \exp\left\{\sum_{i=1}^{\infty}\alpha_{i}\lambda\left(s^i-1\right)\right\}, \quad s \in \mathbb{R},
\label{eqn:G}
\end{equation}
where $\boldsymbol{\alpha}=(\alpha_1,\alpha_2,\ldots) \in \mathbb{R}^{\infty}$ is the probability set describing a discrete random variable $X_i$ with the sum $\sum_{i=1}^{\infty}\alpha_i = 1$. Thus, the DCP random variable $Y$ is denoted by:
$$
Y \sim \text{DCP}(\lambda,\boldsymbol{\alpha}).
$$
The pgf in the Equation (\ref{eqn:G}) is used to study the characteristics of DCP distribution that includes many widely-used models such as Poisson, Hermite \citep{kemp1965some}, Neyman Type A \citep{neyman1939new}, negative binomial and geometric-Poisson \citep{polya1930quelques}. The following subsections provide the key characteristics of a few popular DCP distributions including Poisson, negative binomial and geometric-Poisson distributions. These three distributions refer to three plausible cases of ``no clustering'', ``logarithmic distributed clustering'', ``geometric distributed clustering'' respectively.

\subsubsection{Poisson distribution}
Poisson distribution is the trivial case of DCP distribution where the random variables $X_i$ in the Equation (\ref{eqn:dcp}) is a degenerated distribution with probability mass function (pmf):
$$
p_{X_i}(x_i) =
\begin{cases}
1, & x_i=1\\
0, & \text{otherwise}.
\end{cases}
$$
Therefore, the parameter $\boldsymbol{\alpha}$ is set to a unit vector of $\alpha_1=1$. Thus, the pgf of the Poisson random variable $Y$ is 
$$
G_Y(s) = \exp\left(\lambda(s-1)\right).
$$

\subsubsection{Negative binomial distribution}

Negative binomial distribution is proven to be a DCP distribution with a logarithmically distributed counts of each cluster \citep{quenouille1949relation}. The pmf of the logarithmic distribution is given by:
$$
p_X(x) = \frac{-1}{\log (1-p)} \frac{p^x}{x}, \quad x = 1,2,\ldots,
$$
where $0<p<1$. The pgf of the negative binomial distribution can be derived by considering a composition function of the pgf of Poisson distribution and the pgf of logarithmic distribution \citep{johnson2005univariate}. The composition function of the $G_N(s) = \exp\left(\lambda(s-1)\right)$ and $G_X(s) = \log(1-ps)/\log(1-p)$ is given by:
$$
G_Y(s) = G_N(G_X(s)) = \left(\frac{p}{1-(1-p)s}\right)^{-\frac{\lambda}{\log p}}.
$$
A parameterisation used by \cite{gelman1995bayesian} is adopted in this paper with a pmf:
$$
 p_Y(y) = {y+\kappa-1 \choose y}\left(\frac{\lambda}{\lambda+\kappa}\right)^y\left(\frac{\kappa}{\lambda+\kappa}\right)^\kappa.
$$
The expectation, variance and variance-to-mean ratio are $E[Y]= \lambda$, $Var[Y]=\lambda+\lambda^2/\kappa$ and $D=1+\lambda/\kappa$ respectively. The pgf is given by:
$$
\left(\frac{\kappa}{\lambda(1-s)+\kappa}\right)^\kappa.
$$
The distribution also includes Poisson model as a limiting case, i.e.:
$$
\lim_{\kappa \to \infty} G_Y(s) = \lim_{\kappa \to \infty} \left(1+\frac{\lambda(1-s)}{\kappa}\right)^{-\kappa} = \exp\left\{\lambda(s-1)\right\}.
$$

\subsubsection{Geometric-Poisson distribution}

An alternative approach to account for clustered counts in the Poisson model is to
place a geometric distribution on each cluster count. A geometric-Poisson random variable is defined on a DCP random variable with a zero-truncated geometric random variable within clusters. \cite{ozel2010probability} derive the mass function as:
$$
p_Y(y) =
\begin{cases}
e^{-\lambda}, & y=0\\
e^{-\lambda}\sum^{y}_{k=1}\frac{\lambda^k}{k!}{{y-1}\choose{k-1}}\theta^k(1-\theta)^{y-k}, & y = 1,2,3,\ldots
\end{cases}
$$
where $\lambda>0$, $0<\theta<1$. The expectation, variance and variance-to-mean ratio are $E[Y]=\lambda/\theta$, $Var[Y] = \lambda(2-\theta)/\theta^2$ and $D=(2-\theta)/\theta$ respectively. The pgf is given by
$$
G_Y(s) = \exp\left\{-\lambda\left[1-\frac{(1-\theta)s}{1-\theta s}\right]\right\},
$$
which is clearly a pgf of the DCP distribution in the Equation (\ref{eqn:G}).

\subsection{Compound Poisson regression}
\noindent Consider a regression model with the following covariates:
$$
\begin{array}{rcl}
\log \lambda_i & = & \beta_0 + \beta_1 \log(\text{TG}_i) + \beta_2 \log(|\text{SUP}|_i+0.01) + \beta_3 \text{TCTarget}_i + \\
& &  \beta_4 \log(\text{HomeAvg3}_i+0.01) + \beta_5 \log(\text{AwayAvg3}_i+0.01) + \\ 
& & \beta_6 \log(\text{HomeShotOnGoalAvg3}_i+0.01) +\\
& & \beta_7 \log(\text{AwayShotOnGoalAvg3}_i+0.01),
\end{array}
$$
where $\text{HomeAvg3}_i$ and $\text{AwayAvg3}_i$ are the home and away team's average counts in their previous 3 matches imputed by the overall league average if lack of the previous match information. The covariate $\text{TCTarget}_i$ generated from the target encoding method in the Equation (\ref{eqn:te}) handles the adjusted mean corner count. $\text{HomeShotOnGoalAvg3}_i$ and $\text{AwayShotonGoalAvg3}_i$ are the home and away team's number of shots in their previous 3 matches. If the team has not competed previously, the counts are imputed by the league-dependent value encoded in the Equation (\ref{eqn:te}) and by the grand mean if no league information is available. Adding a small positive constant to a covariate is a popular approach to tackle the log transformation of values that include zero \citep{bellego2021dealing}.

\subsection{Regression on the shape parameters}

On the grounds that the flexibility of Bayesian hierarchical model, an additional regression structure can be added to the negative binomial and geometric-Poisson shape parameters $\kappa_i$ and $\theta_i$. The regression components on the shape parameter are defined as:
$$
\begin{array}{llll}
\text{negative binomial regression:} & \log(\kappa_i)  & = & \alpha_0 + \alpha_1 \log(|\text{SUP}_i|+0.01)\,\\
\text{geometric-Poisson regression:} & \text{logit}(\theta_i)  & = & \alpha_0 + \alpha_1 \log(|\text{SUP}_i|+0.01).
\end{array}
$$
Apart from using the implied goal supremacy considered above as regression covariate, target encoding method can be incorporated in the shape component. The maximum likelihood estimates of the shape parameters for each competition according to the target encoding method in Equation (\ref{eqn:te}) can be used as the covariate. For estimating the dispersion parameter $\kappa$ of negative binomial distribution, \cite{levin1977compound} prove the existence of unique solution provided that $s^2 > \bar{y}$. The same treatment is also applied to the geometric-Poisson model. However, some of the league-dependent shape parameter estimates give large standard errors. The covariate $\log(|\text{SUP}_i|+0.01)$ is a reasonable choice for capturing the heterogeneity of corner counts between matches.

\section{Model performance, diagnostics and betting simulation}

The existence of favourite-longshot bias in the corner O/U, HAD and number of total goals scored O/U markets are examined by a negative mean logarithmic scoring rule \citep{gneiting2007strictly}. The R package \texttt{implied} \citep{implied} offers an implementation of popular margin removal methods to calculate implied probabilities of betting odds namely multiplicative method, odds ratio method, power method and Shin method \citep{shin1993measuring, clarke2017adjusting}. Except the multiplicative method allocates margin proportionally to the odds, all other methods penalise longshot and apply less margin on favourite selections. The corner O/U bet-type is defined as the total number of corners taken being higher or lower than the number specified. Table \ref{tab:margin} shows that the multiplicative method provides optimal predictability on corner O/U and the number of total goals scored O/U markets whilst Shin's method offers the best prediction by discounting longshot probabilities. Therefore, the Shin method and the multiplicative method are used for handling $\textbf{TG}$ and $\textbf{SUP}$ covariates.

\begin{table}[hbt!]
\caption{Performance comparison of margin removal methods}
\begin{tabular}{lrrr}
\hline
 & \multicolumn{3}{c}{Negative mean logarithmic scoring rule} \\                 
Method & Corner O/U & HAD & Total goals O/U \\\hline
Multiplicative & \textbf{14187.07} & $19618.98$ & $\textbf{14465.44}$  \\
Odds ratio & 14209.67 & $19614.28$ & $14522.77$ \\
Power & 14219.82 & $19633.16$ & $14548.68$ \\
Shin & 14207.74 & $\textbf{19612.15}$ & $14517.71$\\
\hline
\end{tabular}
\label{tab:margin}
\end{table}

Five candidate models are considered here, namely na\"{i}ve Poisson, negative binomial and geometric-Poisson regression models; negative binomial model and geometric Poisson models with varying shape parameters. The result of the preliminary fitted models shows that only $\text{TG}, \text{TCTarget}$, $\text{HomeAvg3}$ and $\text{AwayAvg3}$ are significantly associated with the number of the total corner kicks taken. Although non-significant covariates $\text{SUP}$, $\text{HomeShotOnGoalAvg3}$ and $\text{AwayShotOnGoalAvg3}$ with $95\%$ credible interval of covariate estimates covering zero do not provide a good insight to the mean value of corner count. They are able to capture the variation of the distribution’s higher moment hence the predictability. For example, a high absolute value of goal supremacy can be translated to a more unpredictable corner count. The models are fitted by Markov chain Monte Carlo (MCMC) algorithms via No U-turn sampler \citep{hoffman2014no} implemented in Stan language \citep{carpenter2017stan} with 4000 iterations of each. The overall fit of each model is summarised in Table \ref{tab:modelfit}. To check whether all candidate models are well-specified, the Pareto K is examined that all estimates are less than $0.5$. It shows all fitted models have high reliability and exhibit convergence. Leave-one-out expected log pointwise predictive density (elpd\_loo) is used for model comparison \citep{vehtari2017practical}. The geometric-Poisson regression model has the highest elpd\_loo which fits the data better than other models. However, an additional regression component on the shape parameter does not improve the goodness-of-fit of the geometric Poisson regression model since it has an even smaller elpd\_loo whilst the negative binomial regression model with varying shape parameter has a significantly higher elpd\_loo value (the se\_diff between two models is 5.7 which is not shown in the Table \ref{tab:modelfit} than its fixed shape parameter counterpart. 

\begin{table}[hbt!]
\caption{Model performance by the expected log-pointwise predictive density (elpd\_loo) and LOO information criterion (looic) of all candidate models, where p\_loo is the estimated effective number of parameters of the model, se\_diff is the standard error of the difference in the elpd estimate to the one for geometric-Poisson model with varying shape parameter}
\begin{tabular}{lrrr}
\hline
Model & elpd\_loo (criterion) & p\_loo (penalty) &  se\_diff    \\\hline
(A) Geometric-Poisson & \textbf{$-30856.4$} & $10.0$ & $0.0$ \\
(B) Geometric-Poisson + shape reg & $-30856.8$ & $11.5$ & $1.4$\\
(C) Negative binomial + shape reg & $-30858.4$ & $11.6$ & $1.7$\\
(D) Negative binomial & $-30871.4$ & $9.1$ & $5.3$\\
(E) Poisson & $-30877.2$ & $10.1$ & $6.9$\\
 \hline
\end{tabular}
\label{tab:modelfit}
\end{table}

Table \ref{tab:parameters2} shows that implied number of total goals scored has a positive effect on the number of corner kicks. For each of the expected goal, the expected corner count is increased by $16.70\%$ for the model A. Previous corner kicks taken and conceded by the same home and away teams also play a crucial role. The league average corner counts exerts a negative effect on the number of corner kicks taken. It can be seen as an adjustment to a stronger effect of previous corner kicks taken and conceded. The median of the posterior distribution of model A's shape parameters $\theta$ is $0.9577$. It indicates that a single corner induces additional 0.0423 corners and also implies that the variance-to-mean ratio has an estimate of 1.0883. 

\begin{table}[hbt!]
\caption{Parameter estimates of all candidate models A-E}
\begin{tabular}{lccc}
\hline
Parameter & \multicolumn{3}{c}{Model} \\
& (A) Geometric-Poisson & (B) GP + shape & (C) NB + shape \\\hline
$\beta_0$ & $0.711 (0.0219,1.4154)$ & $0.788 (0.0938,1.4654)$ & $0.7782 (0.0562,1.5142)$ \\ 
$\beta_1$ & $0.1545 (0.1046,0.2069)$ & $0.1688 (0.1216,0.2231)$ & $0.1573 (0.099,0.2109)$ \\ 
$\beta_2$ & $-0.001 (-0.007,0.0049)$ & $-0.0018 (-0.0072,0.0037)$ & $-0.001 (-0.0068,0.0047)$ \\ 
$\beta_3$ & $-0.6168 (-0.9374,-0.2813)$ & $-0.6588 (-1.0025,-0.3218)$ & $-0.6206 (-0.9676,-0.2577)$ \\ 
$\beta_4$ & $0.5957 (0.53,0.6625)$ & $0.6092 (0.5436,0.6725)$ & $0.5957 (0.5326,0.6628)$ \\ 
$\beta_5$ & $0.6052 (0.5405,0.6652)$ & $0.609 (0.5417,0.6754)$ & $0.5996 (0.537,0.6659)$ \\ 
$\beta_6$ & $-0.0024 (-0.0084,0.0033)$ & $-0.0029 (-0.0086,0.0033)$ & $-0.0022 (-0.0084,0.0037)$ \\ 
$\beta_7$ & $-0.0019 (-0.0088,0.0043)$ & $-0.0025 (-0.0088,0.0042)$ & $-0.002 (-0.009,0.0039)$ \\ 
$\theta$ & $0.9577 (0.9441, 0.9705)$ \\ 
$\alpha_0$ && $3.7470 (2.9766,4.8491)$ & $4.7198 (4.3913,5.1212)$\\
$\alpha_1$ && $-0.1978 (-0.7523,0.1141)$ & $36.3012 (-12.0218,91.3719)$\\
 \hline
 & \multicolumn{3}{c}{Model} \\
& (D) Negative binomial & (E) Poisson  \\\hline
$\beta_0$ & $0.7575 (0.0924,1.4591)$ & $0.7438 (0.0774,1.4203)$ \\ 
  $\beta_1$ & $0.1559 (0.1045,0.2028)$ & $0.156 (0.1032,0.2041)$ \\ 
  $\beta_2$ & $-0.001 (-0.0066,0.0049)$ & $-0.0009 (-0.0061,0.0045)$ \\ 
 $\beta_3$ & $-0.607 (-0.9541,-0.2698)$ & $-0.6048 (-0.9345,-0.2794)$ \\ 
  $\beta_4$ & $0.5931 (0.5273,0.6607)$ & $0.5959 (0.532,0.6686)$ \\ 
  $\beta_5$ & $0.5974 (0.5343,0.6664)$ & $0.6043 (0.5434,0.6659)$ \\ 
  $\beta_6$ & $-0.0024 (-0.0082,0.0043)$ & $-0.0023 (-0.0081,0.004)$ \\ 
  $\beta_7$ & $-0.0019 (-0.0089,0.0044)$ & $-0.0021 (-0.0088,0.0039)$ \\ 
  $\kappa$ & $59.8998 (52.2233,68.2215)$\\
  \hline
\end{tabular}
\label{tab:parameters2}
\end{table} 

\begin{figure}[hbt!]
  \noindent\includegraphics[width=30pc,angle=0]{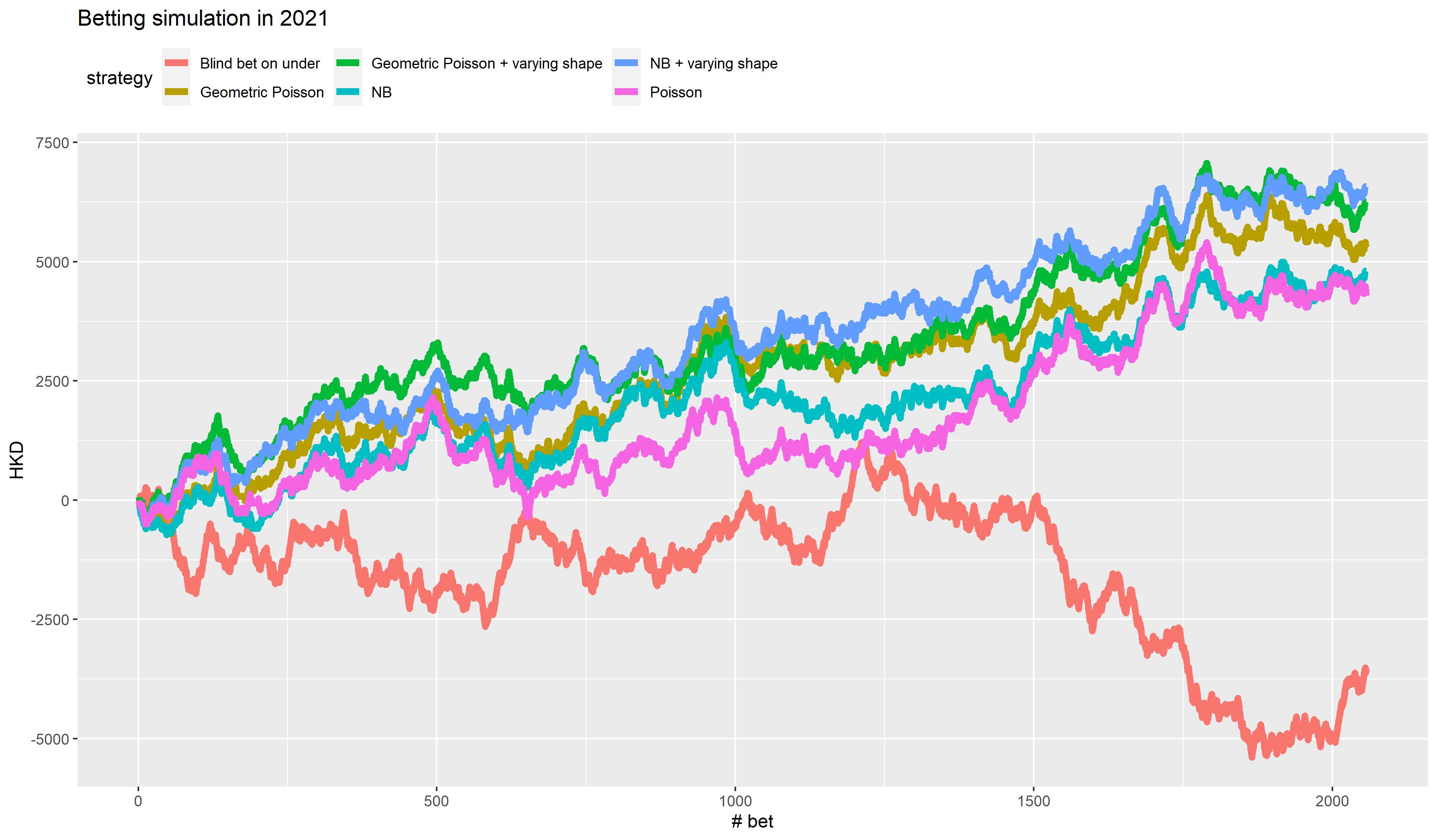}\\
  \caption{Cumulative profits for 2057 matches offered by the HKJC in 2021 from 1st January to 10th June in 2021.}
  \label{fig:sim}
\end{figure}

The betting simulation is performed for 2057 matches offered by the HKJC in 2021 from 1st January to 10th June in 2021 (Figure \ref{fig:sim}). Although average overall margin taken by the operator is 7.61\% (payout rate 92.38\%). However, the payout rate on the ``over'' selection is 85.71\% whilst the ``under'' selection one is 98.24\%. It indicates that, empirically, the market has a bias in favour of the ``under'' selection.

A bet of $100$ dollars is placed on each of the selection when the model expected value is positive. With the five candidate models considered in Table \ref{tab:modelprofit} along with a blind betting strategy on all ``under'' outcomes, cumulative return is calculated for the simulated period. Sharpe ratio \citep{sharpe1966mutual} is calculated from the ratio of aggregated daily profit and its sample standard deviation annualised by the square root of the total 364 trade days \citep{sharpe1966mutual,lo2002statistics}. The negative binomial regression model with match varying shape parameter outperforms other candidates and gives the best final profit of $\$6578$ and Sharpe ratio $3.065$. The Table (\ref{tab:parameters2}) shows that the location level regression parameter estimates ($\beta_0 - \beta_7$) of the negative binomial model with match varying parameter (model C) are almost identical to those in model (A). The estimate of the parameter $\alpha_1$ indicates that the variance-to-mean ratio $D$ has a weak but notable relationship with the magnitude of the implied goal supremacy $\textbf{SUP}_i$. 

\begin{table}[hbt!]
\caption{Out-of-sample betting simulation for the period between 1st Jan 2021 to 10th June 2021 ranked by its Sharpe ratio}
\begin{tabular}{lrrrr}
\hline
Model & \# of bets & profit $\$$ & profit $\%$& Sharpe ratio    \\\hline
(C) Negative binomial + shape reg & $1137$ & $6578$ & $5.785\%$ &  $3.065$\\
(B) Geometric-Poisson + shape reg & $1125$ & $6256$ & $5.561\%$ & $2.922$\\
(A) Geometric-Poisson & $1123$ & $5308$ & $4.727\%$ & $2.427$\\
(D) Negative binomial & $1085$ & $4805$ & $4.429\%$ & $2.263$\\
(E) Poisson & $1156$ & $4283$ & $3.705\%$ & $1.519$\\
Blind bet on under & $2056$ & $-3530$ & -1.717\%&  $-1.267$\\
 \hline
\end{tabular}
\label{tab:modelprofit}
\end{table}


\section{Discussion}

This article presents a Bayesian overdispersed Poisson regression framework to forecast future corner counts leading to a profitable forecasting model. The model incorporates multiple sources of match and market information including previous team's corner counts, cross-market information and competition-level statistics. This work could be also extended to accommodate underdispersed count through the Convey-Maxwell-Poisson (CMP) distribution \citep{conway1962queuing}. The probability mass function is given by:
$$
p_Y(y) = \frac{\lambda^y}{\left(y!\right)^\nu}\frac{1}{Z(\lambda,\nu)},
$$
where $Z(\lambda,\nu)=\sum^\infty_{j=0}\frac{\lambda^i}{(j!)^\nu}$. When $\nu=1$, the CMP distribution becomes Poisson distribution; when $\nu=0$, it becomes geometric distribution. A recent result in \cite{geng2022conway} shows that a CMP random variable is infinitely divisible if it is Poisson or geometric. This theoretical result corroborates that the CMP distribution does not fall into the DCP distribution family. Exploiting the property of exponential family, a generalised linear model (GLM) for CMP distribution with two link functions on the location and shape parameters can be applied (e.g.: \citealt{guikema2008flexible}), the interpretation is different from the framework introduced in this paper.

Our study shows that the favourite long-shot bias in the HAD odds is captured adequately by the Shin's method. Except the HAD market, the implied probability of the total goals scored and corner O/U markets can be easily calculated through handling the margin weights proportional to the original odds with the multiplicative method. The market efficiency of the corner O/U markets is also biased in favour of the ``under'' selection. The asymmetry toward the ``under'' selection is understood as the market overvalues the ``over'' selection. The margin removal methods utilised in this study is unable to capture this kind of asymmetry. Along with the favourite-longshot bias studied previously, the O/U asymmetry is an area one should work on.

Our compound Poisson framework allows
a complex model structure with multiple regression equations in the same model. With the application of MCMC algorithms in Stan language, the model development process can be seen as iterative and continuous under the hierarchical Bayesian modelling paradigm. Its flexibility is already proven in other research areas such as epidemiology (for example, \nocite{yip2022spatio}Yip et al., 2022). A possible extension of this model is to expand the regression component with a multi-level hierarchical structure to adjust the estimates of data-rich competitions:
\begin{displaymath}
\begin{array}{lll}
\log\lambda_{i} & = & \log\lambda_{2i} + \boldX'_{1i}\boldsymbol{\alpha}\\
\log\lambda_{2i} & = & \boldX'_{2i}\boldsymbol{\beta},
\end{array}
\end{displaymath}
where the term $X_{2i}\boldsymbol{\beta}$ is the regression component on covariates of all competitions and $X_{1i}\boldsymbol{\alpha}$ is the regression component of data-rich competitions such as inclusion of zonal and player-based statistics. These information can be obtained from OPTA sportsdata \citep{liu2013inter} takes into account of on-pitch events and player actions with spatial locations. From the betting simulation, some competitions with a large fan base such as English Premiere League and European Champions League are recorded negative profits. Further modelling on the data-rich elite competitions will need to be developed and experimented.


\section*{Acknowledgements}
\noindent We thank Jonas Christoffer Lindstrøm to provide helpful supports in using the \texttt{implied} package in R.

\newpage
\clearpage
\bibliography{mybibfile}

\end{document}